\documentclass[times,twocolumn,final,nopreprintline]{elsarticle}
\usepackage{framed,multirow}

\usepackage{amssymb}
\usepackage{latexsym}

\usepackage{url}
\usepackage{xcolor}
\usepackage{svg}

\usepackage{hyperref}
\usepackage{amsmath} 

\DeclareMathOperator*{\argmin}{\arg\!\min}

\definecolor{newcolor}{rgb}{.8,.349,.1}

\begin{document}

% \verso{Noga Kertes \textit{et~al.}}

\begin{frontmatter}

\title{IVIM-Morph: Motion-compensated quantitative Intra-voxel Incoherent Motion (IVIM) analysis for functional fetal lung maturity assessment from diffusion-weighted MRI data
%\tnoteref{tnote1}
}%
%\tnotetext[tnote1]{This is an example for title footnote coding.}
% \author[1]{Noga \snm{Kertes}\fnref{fn1}}
% \author[1]{Yael \snm{Zaffrani-Reznikov}\fnref{fn1}}
% \author[2]{Onur \snm{Afacan}}

% \author[2]{Sila \snm{Kurugol}}
% %% Third author's email
% %\ead{author3@author.com}
% \author[2]{Simon K. \snm{Warfield}}
% \author[1]{Moti \snm{Freiman}\corref{cor1}}

\author[1]{Noga Kertes\fnref{fn1}}
\author[1]{Yael Zaffrani-Reznikov\fnref{fn1}}
\author[2]{Onur Afacan}

\author[2]{Sila Kurugol}
%% Third author's email
%\ead{author3@author.com}
\author[2]{Simon K. Warfield}
\author[1]{Moti Freiman\corref{cor1}}

\fntext[fn1]{Equal contribution.}
\cortext[cor1]{Corresponding author: Moti Freiman,   Tel.: +972-77-887-4147; 
  Email: \href{mailto:moti.freiman@technion.ac.il} {moti.freiman@technion.ac.il}}
  
\address[1]{Faculty of Biomedical Engineering, Technion, Haifa, Israel}
\address[2]{Boston Children's Hospital, Boston, MA, USA}

%\received{1 May 2013}
%\finalform{10 May 2013}
%\accepted{13 May 2013}
%\availableonline{15 May 2013}
%\communicated{S. Sarkar}

\begin{abstract}
%%%
Quantitative analysis of pseudo-diffusion in diffusion-weighted magnetic resonance imaging (DWI) data shows potential for assessing fetal lung maturation and generating valuable imaging biomarkers. Yet, the clinical utility of DWI data is hindered by unavoidable fetal motion during acquisition. We present IVIM-morph, a self-supervised deep neural network model for motion-corrected quantitative analysis of DWI data using the Intra-voxel Incoherent Motion (IVIM) model. IVIM-morph combines two sub-networks, a registration sub-network, and an IVIM model fitting sub-network, enabling simultaneous estimation of IVIM model parameters and motion. To promote physically plausible image registration, we introduce a biophysically informed loss function that effectively balances registration and model-fitting quality. We validated the efficacy of IVIM-morph by establishing a correlation between the predicted IVIM model parameters of the lung and gestational age (GA) using fetal DWI data of 39 subjects. Our approach was compared against six baseline methods: 1) no motion compensation, 2) affine registration of all DWI images to the initial image, 3) deformable registration of all DWI images to the initial image, 4) deformable registration of each DWI image to its preceding image in the sequence, 5) iterative deformable motion compensation combined with IVIM model parameter estimation, and 6) self-supervised deep-learning-based deformable registration. 
IVIM-morph exhibited a notably improved correlation with gestational age (GA) when performing in-vivo quantitative analysis of fetal lung DWI data during the canalicular phase. Specifically, over 2 test groups of cases, it achieved an $R^2_{f}$ of $0.44$ and $0.52$, outperforming the values of $0.27$ and $0.25$, $0.25$ and $0.00$, $0.00$ and $0.00$, $0.38$ and $0.00$, and $0.07$ and $0.14$ obtained by other methods.
IVIM-morph shows potential in developing valuable biomarkers for non-invasive assessment of fetal lung maturity with DWI data. Moreover, its adaptability opens the door to potential applications in other clinical contexts where motion compensation is essential for quantitative DWI analysis. The IVIM-morph code is readily available at:\url{https://github.com/TechnionComputationalMRILab/qDWI-Morph}. 
%%%%
\end{abstract}

% \begin{keyword}
% %% MSC codes here, in the form: \MSC code \sep code
% %% or \MSC[2008] code \sep code (2000 is the default)
% %\MSC 41A05\sep 41A10\sep 65D05\sep 65D17
% %% Keywords
% \KWD Quantitative Diffusion-Weighted MRI \sep Fetal imaging \sep  Motion compensation
% \end{keyword}

\end{frontmatter}

%\linenumbers

%% main text
% \label{Introduction}
\section{Introduction}

Congenital pulmonary hypoplasia (PH) is a congenital abnormality marked by insufficient growth of the lung parenchyma \cite{lakshminrusimha2015persistent}. This condition can result in significant and potentially fatal physiological impairments, including respiratory distress syndrome and transient tachypnea of the newborn \cite{ahmed2021fetal}. Approximately 10-15\% of newborn deaths are caused by PH \cite{kumar2007congenital}.

Currently, the methods employed for the antenatal diagnosis of pulmonary hypoplasia include amniocentesis, prenatal ultrasound (US), and MRI. Amniocentesis involves extracting a small volume of amniotic fluid to assess surfactant protein levels, which are considered an indicator of fetal lung maturity \cite{rome1975benefits}. Prenatal ultrasound is a widely utilized technique for evaluating fetal lung maturity. However, it is primarily used to assess fluid parameters \cite{beck2015assessment} or to estimate fetal lung volume \cite{Moeglin2005FetalUltrasound}. Similarly, fetal lung MRI can estimate fetal lung volume \cite{ward2006fetal}. Nonetheless, these modalities do not provide sufficient insight into lung function and are, therefore, suboptimal for assessing functional fetal lung maturity and PH \cite{avena2022assessment}.

Diffusion-weighted MRI (DWI) is a non-invasive imaging modality that is highly sensitive to the random motion of water molecules, which is primarily due to the water's thermal energy. In living tissues, the motion of water molecules is influenced and restricted by interactions with cell membranes and macromolecules. Moreover, the motion of water molecules is more confined in tissues with higher cellular density, while the motion of water molecules is less restricted in areas of low cellularity. In DWI, the random displacement of individual water molecules leads to signal attenuation when magnetic field encoding gradient pulses are applied at varying magnitudes and durations known as the ``b-value'' \cite{koh2007diffusion}.

Quantitative analysis of DWI (qDWI) using multi-compartment signal decay models such as the Intravoxel Incoherent Motion (IVIM) Model \cite{iima2016clinical}, can provide a separate assessment of diffusion and pseudo-diffusion in tissue. This approach allows for more precise imaging biomarkers that capture the key characteristics of functional lung maturity and PH such as the formation of a dense capillary network, an increase in pulmonary blood flow, a reduction in extracellular space, and an increase in tissue perfusion \cite{Ercolani2021IntraVoxelMaturation, korngut2022super,jakab2017intra}.

\begin{figure*}[t!]
\centering
\includegraphics[width = 150mm]{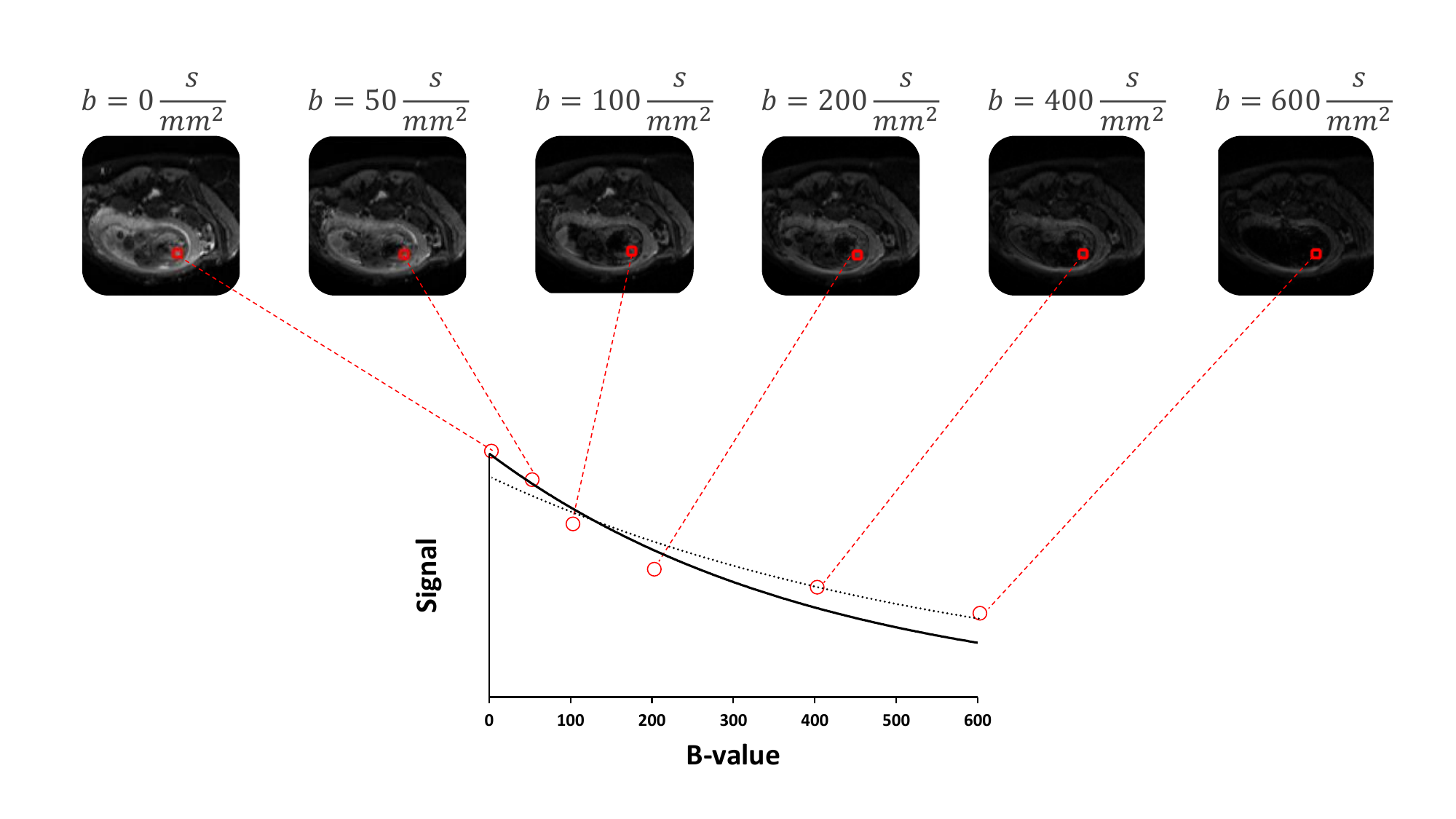}
\caption{Fetal DWI data acquired with varying b-values. Fetal motion causes the observed signal (red circles) to deviate from the expected signal decay model (solid line). 
Fitting the model to the observed signal without accounting for motion may lead to an incorrect estimate of the model parameters (dashed line).}
\label{fig: data_example}
\end{figure*}

However, the inevitable motion of the fetus during lengthy DWI acquisitions generally leads to inaccurate and unreliable quantitative analysis of diffusion and pseudo-diffusion, which effectively renders these imaging biomarkers of little utility in assessing functional lung maturity and PH in the clinical setting. Figure~\ref{fig: data_example} illustrates the deviations of the observed DWI signal acquired at different b-values from the signal decay model.
For instance, \cite{Afacan2016FetalAge} reported that nearly 40\% (26 out of 65 cases) in their study cohort had severe motion artifacts, which essentially prevented the functional assessment of lung maturity with DWI. Thus, there is a critical need to develop methods for qDWI analysis that are robust to the presence of inter-volume motion in fetal DWI data.

Image registration algorithms have been previously used to address inter-volume motion before qDWI analysis. For instance, Guyader et al. \cite{guyader2015influence} demonstrated improved accuracy and reliability in apparent diffusion coefficient (ADC) qDWI analysis of abdominal organs when employing initial motion correction, as opposed to qDWI analysis conducted without any motion correction. However, the registration of DWI images obtained using varying b-values may lead to suboptimal accuracy owing to differences in image contrast caused by varying sensitivities to diffusion and pseudo-diffusion effects. Registration of high b-value images, which have a low signal-to-noise ratio (SNR) by nature, also poses a significant challenge. Moreover, it is worth noting that optimization processes typically optimize loss functions tied to pairwise metrics, such as Dice similarity or intensity dissimilarity. These metrics, inherently designed for pairwise comparisons, possess limitations in their ability to comprehensively address motion across the entire set of DWI images concurrently. Furthermore, their primary focus tends to be on aligning image edges, rather than ensuring precise alignment of the observed signal decay within regions of interest with the signal decay model.

In the context of abdominal imaging, \cite{Kurugol2017Motion-robustEstimation} introduced an iterative motion correction model to address the differences in image contrast in the DWI images by registering images and estimating parameters with the IVIM model. Similarly, \cite{sanz2018robust} simultaneously compensates for motion and performs qDWI analysis using a mono-exponential signal decay model. However, these techniques involve an iterative application of the image registration and model fitting steps. Unfortunately, the iterative process may lead to suboptimal results due to convergence to local minima. Additionally, the computational demand and processing time associated with such methods renders them impractical for clinical use with large datasets. Recently, \cite{kornaropoulos2022joint} presented a novel approach for joint motion correction and quantitative analysis of prostate DWI data using the mono-exponential signal decay model. Specifically, they used a Markov-Random-Field (MRF) technique to simultaneously optimize a motion correction and qDWI model fitting problem. However, this method necessitates the discretization of both domains, and it is computationally intensive. Thus, clinically viable methods for qDWI analysis that are resilient to inter-volume motion artifacts are urgently needed.

Recently, within the field of anatomical fetal MRI reconstruction, several deep-learning models have emerged. Cordero-Grande et al. \cite{cordero2022fetal} introduced a deep generative prior model, while Xu et al. \cite{xu2023nesvor} adopted an implicit neural representation approach to achieve motion-robust volumetric reconstruction of anatomical fetal MRI data. In a related context, Davidson et al. \cite{davidson2022motion} utilized slice-to-volume deformable image registration to extract reliable 3D measurements of fetal lung volume from fetal MRI data. However, it's worth noting that all these studies primarily focus on slice-to-volume registration within the scope of a single-volume anatomical MRI, without considering potential motion between the different volumes required for quantitative DWI analysis in functional fetal lung maturation assessment.

In this study, we tackle this challenge with the introduction of a self-supervised Deep Neural Network (DNN) framework named ``IVIM-Morph.'' This approach addresses simultaneous motion compensation and bi-exponential IVIM model parameter estimation. Our model comprises two key sub-networks: the first focuses on estimating deformation fields for motion correction, while the second predicts IVIM model parameters based on the motion-corrected data. To ensure the consistency of DWI signal decay with the IVIM model, we introduce an innovative, physics-based loss function. This loss function penalizes signal decays that deviate from the expected IVIM model behavior, thus maintaining physical plausibility. Importantly, our DNN model significantly reduces computation time compared to conventional methods.

% To evaluate the benefits of our approach in obtaining more accurate and reliable IVIM model parameter estimates in the presence of motion, we conducted numerical simulations and evaluated the IVIM signal decay model fit quality using  13 clinical, fetal DWI datasets without observed motion. We have shown that the IVIM-Morph not only effectively compensates for motion at various motion levels but also prevents the illusion of motion in cases \underline{without} any actual motion.
We assessed the anatomical registration accuracy of our method by manually delineating one lung in each DW image from 5 cases with severe motion artifacts and 5 cases with moderate motion artifacts. We then evaluated the alignment of the masks before and after registration using IVIM-Morph in comparison to various registration techniques. Additionally, we have showcased the clinical importance of using IVIM-Morph for reliable IVIM parameter estimation in the presence of motion by illustrating its capability to enhance the correlation between the predicted perfusion fraction parameter ($f$) in the fetal lung and gestational age (GA) through an analysis of 39 clinical fetal DWI datasets.

Our study delivers the following key contributions:
\begin{itemize}
    \item We offer a self-supervised deep-learning-based mathematical framework for concurrently estimating motion correction and signal decay model parameters.

    \item We introduce an innovative registration loss function, guaranteeing physically sound deformation fields that align with the signal decay model.

    \item We present a comprehensive assessment of our approach, encompassing registration accuracy and its clinical applications in evaluating fetal lung functional maturity.

    \item We will make our code repository, facilitating motion-compensated IVIM analysis of DWI data, accessible to the public.

\end{itemize}

\section{Background}

The bi-exponential IVIM model describes the DWI signal attenuation at a particular voxel relative to the baseline signal as a function of the b-value used during the acquisition \cite{iima2016clinical}:
\begin{equation}
 f_{IVIM}(b_i,f,D^*,D)=S_{0}\left(f\cdot e^{-b_i (D+D^*)} +(1-f)\cdot e^{-b_i \cdot D}\right)
\label{eq:IVIM}
\end{equation}

where $S_{0}$ is the baseline signal obtained without applying any diffusion-synthesized gradients; $D$ is the diffusion coefficient; $D^*$ is the pseudo‐diffusion coefficient; $b_i$ is the b‐value used during the acquisition; and $f$ is for the perfusion fraction \cite{federau2017intravoxel,iima2021perfusion}.

The estimation of the IVIM model parameters from the DWI data acquired with multiple b-values $B=\left\{b_i\right\}_{i=0}^{N}$, is commonly done by solving the following least-squares problem:
\begin{equation}
\widehat{D},\widehat{D^*},\widehat{f} =  \argmin_{D,D^*,f} \sum_{b_i \in B}\left\|  
S(b_i)-f_{IVIM}(b_i,f,D^*,D) \right\|^2 
 \label{eq:ivim_ls}
 \end{equation}
Supplementary regularization terms are frequently incorporated to enhance estimation robustness in the presence of noise and improve clinical diagnostic accuracy \cite{freiman2013reliable,orton2014improved,spinner2021bayesian,vidic2019accuracy}.

In the past few years, state-of-the-art, DNN-based methods were introduced for IVIM parameter estimation. Bertleff et al. \cite{bertleff2017diffusion} demonstrated the ability of supervised DNN to predict the IVIM model parameters from low SNR DWI data. Barbieri et al. \cite{barbieri2020deep} proposed an unsupervised, physics-informed DNN (IVIM-NET) with results comparable to Bayesian methods with further optimizations by Kaandorp et al. \cite{kaandorp2021improved} (IVIM-NET$_{\textrm{optim}}$). Zhang et al. \cite{zhang2019implicit} used a multi-layer perceptron with an amortized Gaussian posterior to estimate the IVIM model parameters from fetal lung DWI data. Recently, Vasylechko et al.  \cite{vasylechko2022self} used unsupervised convolutional neural networks (CNN) to improve the reliability of IVIM parameter estimates by leveraging spatial correlations in the data. 

Nevertheless, all these algorithms presuppose spatial alignment among the different b-value images, rendering them unsuitable for direct application in estimating IVIM model parameters for fetal DWI data, given the inevitable fetal motion during acquisition \cite{Afacan2016FetalAge}.

\section{Method}

We address the challenge of estimating the IVIM model parameters while compensating for motion artifacts by presenting a self-supervised DNN-based framework for simultaneous motion compensation and IVIM model parameters estimation. 
Specifically, we aim to find the optimal values for the IVIM model parameters $D$, $D^*$, and $f$, as well as the set of transformations $\Phi=\left\{\phi_i\right\}_{i=0}^N$ that align the observed DWI signals to the model predictions. The joint optimization problem is formulated as follows:
\begin{equation}
\widehat{\Phi},\widehat{D},\widehat{D^*},\widehat{f}=\argmin_{\Phi,D,D^*,f}
\sum_{i=0}^N\left\|\phi_i \circ S(b_i)-f_{IVIM}(b_i,f,D^*,D) \right\|^2 
 \label{eq:problem_formulation_base}
 \end{equation}
where $\phi_i$ represents the spatial transformation that aligns the $i$-th DWI signal to the reference signal (i.e., the model prediction at the $i$-th b-value).

However, direct optimization of this equation is challenging due to the huge number of unknowns associated with the combination of the bi-exponential IVIM signal decay model and the set of free-form deformations.

Instead, we frame this optimization problem as an estimation of the weights of a DNN that predicts IVIM model parameters and spatial transformations from the observed DWI signals and b-values. Specifically, we will minimize the following objective function:
\begin{equation}
 \widehat{\Theta} =  \argmin_{\Theta} \sum_{i=0}^N\left\|f_{\Theta}[0]_i \circ S(b_i)-f_{IVIM}(b_i,f_{\Theta}[1]) \right\|^2 
 \label{eq:problem_formulation_DNN}
 \end{equation}
where $\Theta$ are the parameters of the neural network; $f_{\Theta}$ is the forward pass of the DNN function; and $f_{\Theta}[0]$ and $f_{\Theta}[1]$ represent the DNN outputs corresponding to the set of spatial transformations $\Phi$ and the IVIM parameters $D^*$, $f$, and $D$, respectively.

\subsection{Network architecture}
\begin{figure*}[t!]
\centering
\includegraphics[width=0.98\textwidth]{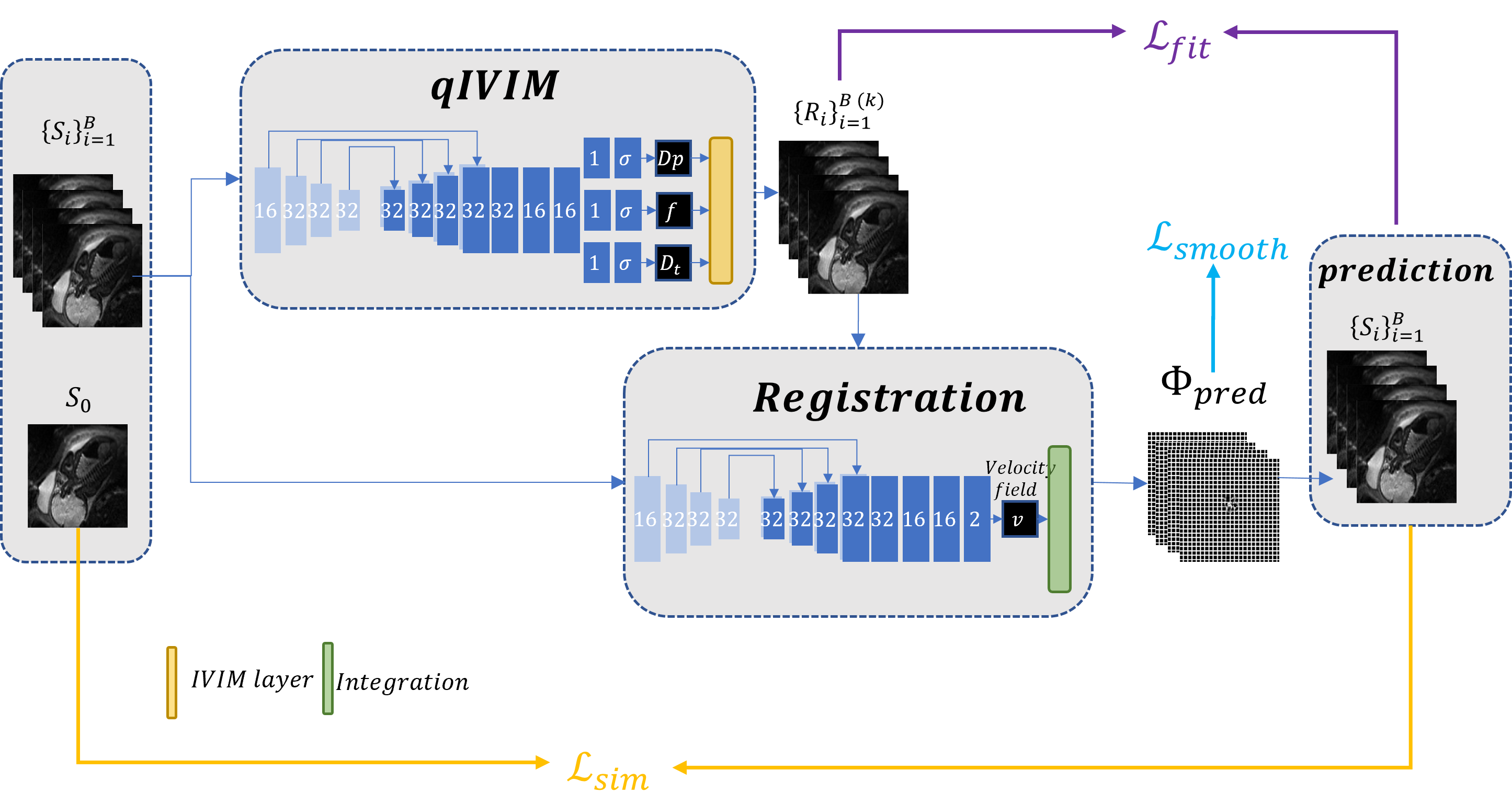}
\caption{The architecture of the IVIM-Morph network, comprises two sub-networks: a quantitative IVIM (qIVIM), a convolutional neural network (CNN), and an image registration sub-network. The qIVIM-CNN sub-network extracts IVIM parameters from the DWI data, while the image registration sub-network aligns each b-value image with the corresponding image reconstructed by the IVIM parameters.}
\label{fig:Arictecture}
\end{figure*}

Figure~\ref{fig:Arictecture} presents the architecture of 
our IVIM-Morph network. It is comprised of two components: a quantitative IVIM (qIVIM-CNN) prediction network and an image registration network. The qIVIM-CNN network is responsible for predicting the IVIM parameters from the DWI data, while the image registration network is responsible for predicting the set of transformations that align each DWI image with the corresponding image reconstructed from the IVIM parameters with Eq.~\ref{eq:IVIM}. We describe each component in the following sections.

\subsubsection{Quantitative IVIM model fitting sub-network}
The qIVIM-CNN is based on a Unet-like architecture \cite{ronneberger2015u} with three parallel decoders, one for each of the IVIM parameters ($D, D^*$, and $f$) \cite{vasylechko2022self}. To ensure physically plausible IVIM model parameter estimates, we used a Sigmoid activation function at the output of each decoder \cite{kaandorp2021improved}:
\begin{equation}
P = P_{min} + Sigmoid(X_P) \times (P_{max}-P_{min})
\label{eq:sigmoid}
\end{equation}
where $P$ denotes any of the IVIM model parameters ($D, D^*, f$); $P_{min}$ and $P_{max}$ are the prior bounds on the parameter, and $X_P$ is the output parameter map from the corresponding Unet decoder.
Table~\ref{Table: parameters range} provides a summary of the boundaries used to constrain the estimates of IVIM model parameters. These boundaries were determined through an IVIM analysis of cases from our database without any significant motion observed using a segmented-least-squares approach \cite{gurney2018comparison} followed by non-linear trust-region-reflective optimization (SLS-TRF) \cite{GradientMethodforLargeScale}.
\begin{table}[t!]
\caption{Prior bounds on the IVIM parameters}
\label{Table: parameters range}
\centering
\begin{tabular}
{ |p{1.5cm}|p{1cm}|p{1cm}|p{1cm}|  }
\hline
Parameter & $D (\frac{mm^2}{sec})$ & f (\%) & $D^* (\frac{mm^2}{sec}$)\\
\hline
minimum & 0.0003 & 7 & 0.006 \\
\hline
maximum & 0.0032 & 50 & 0.15 \\
\hline
\end{tabular}
\end{table}

\subsubsection{Registration sub-network}
We have utilized the Voxel-Morph DNN architecture, renowned for deformable medical image registration \cite{balakrishnan2019voxelmorph,dalca2019varreg, dalca2018varreg}, as the foundation for our image registration sub-network. It predicts the deformation fields ($\Phi=\left\{\phi_{i}\right\}_{i=1}^{N}$) between the acquired DWI data ($\left\{S(b_i)\right\}_{i=1}^{N}$) to the corresponding model-reconstructed images ($\left\{ R(b_i) \right\}_{i=1}^{N}$). The registration is performed between corresponding acquired b-values images and predicted model images such that the moving image $S(b_i)$ is registered to the fixed image $R(b_i)$.
Through the registration of original images to those reconstructed by a predicted model, the network can successfully mitigate variations in the contrast between b-value images. Further, this allows for the utilization of physical prior knowledge via the IVIM model, which characterizes expected signal decay behavior. 

\subsection{Bio-physically-informed loss function}
We introduce an innovative loss function comprising a weighted combination of the following three terms:
\begin{equation}
\label{eq:loss}
\mathcal{L} = \alpha_1 \mathcal{L}_{fit} + \alpha_2 \mathcal{L}_{smooth}  +\alpha_3 \mathcal{L}_{sim}
\end{equation}

% This loss function assesses the concordance between images generated using the predicted signal decay model parameters and the motion-corrected images, while also incorporating a term for deformation field smoothness. Given that our model predicts only the signal decay parameters ($D$, $D^{*}$, and $f$), the baseline signal map $S_0$ remains unchanged. Consequently, we introduced a third term to promote similarity between each deformed image and the baseline image $S_0$.

The model fitting loss ($\mathcal{L}_{fit}$) drives the IVIM-Morph to generate deformation fields that minimize the disparity between acquired images and model-generated images (Eq.~\ref{eq:IVIM}). This guarantees a physically plausible representation of signal decay across the b-value axis, leading to enhanced precision in IVIM parameter maps. Specifically, $\mathcal{L}_{fit}$ is a weighted version of the standard error of the regression (WSER) between the models' prediction of the Diffusion-Weighted images ($\left\{ R(b_i) \right\}_{i=1}^{N}$) and the corresponding deformed images ($\left\{ \phi_i \circ S(b_i)\right\}_{i=1}^{N}$) that accounts for potential bias in high b-value images that have a low signal as follows:
\begin{equation}
\label{eq:WSER}
\mathcal{L}_{fit} = \sqrt{\frac{\sum_{i=1}^{N}w_i\cdot\left(\phi_i \circ S(b_i)-R(b_i)\right)^2}{\sum_{i=1}^{N}w_i}\cdot\frac{1}{N-p-1}}
\end{equation}
The parameter $N$ is the number of observations, which is equal to the number of b-values used during the scan. The parameter $p$ denotes the number of unknowns in the model, which is three in our case ($D$, $D^*$, and $f$), and the weight ($w_i$) is defined as $log(b_i+1)+1$. We normalized the WSER by the average intensity value of the motion-compensated set.

The smoothness term ($\mathcal{L}_{smooth}$) encourages the creation of deformations that are both realistic and invertible. This loss penalizes for a large $L_1$ norm of the gradients of the velocity field $u$ \cite{balakrishnan2019voxelmorph}:
\begin{equation}
 \mathcal{L}_{smooth}(\Phi) = \frac{1}{N} \frac{1}{|\Omega|}\sum_{i=1}^{N} \left \|\nabla u_i \right \|^2
\end{equation}
where $\Omega$ is the image spatial domain.

Lastly, the registration loss ($\mathcal{L}_{sim}$) promotes the alignment of deformation fields, aligning the Diffusion-Weighted images ($S(b_i),i \in \left\{1,..., N\right\}$) with the baseline image $S_0$, by calculating their local normalized cross-correlation (NCC) \cite{balakrishnan2019voxelmorph} 
\begin{equation}
\mathcal{L}_{sim}(S(b_0),\Phi \circ S) = \frac{1}{N} \frac{1}{|\Omega|}\sum_{i=1}^{N}NCC\left(S(b_0), \phi_i\circ S(b_i) \right)
\end{equation}

\subsection{Implementation details}
We implemented our models on Visual Studio Code 1.79.2, Python 3.8.12 with PyTorch 1.13.0, and CUDA 11.8. We applied our suggested methods with a batch size of one, meaning that each batch consists of data from one patient with size: $n_b \times n_x \times n_y$, where $n_b$ is the number of b-values used for scanning the patient, and $n_x \times n_y$ is the image shape. We used an Adam optimizer with an initial learning rate of $10^{-3}$ with a ``reduce on plateau'' learning rate decay scheduler. All the calculations in this study were carried out on a Linux machine equipped with a Tesla V100-PCIE-32GB GPU. The CPU in use was an Intel(R) Xeon(R) Gold 6230 CPU, operating at 2.10GHz.

%We used a grid search approach for hyper-parameter  optimization and set the weights of our loss function (Eq.~\ref{eq:loss}) as follows: $\alpha_1$ is set to 0.8, $\alpha_2$ is 0.015 and $\alpha_3$ is equal to 1.

\section{Experiments}

\subsection{Data}
We used a legacy fetal DWI dataset in this study \cite{Afacan2016FetalAge}. DWI data were acquired on a Siemens 3T Skyra scanner using an 18-channel body matrix coil. The imaging technique used was a multi-slice, single-shot echo-planar imaging (EPI) sequence for obtaining diffusion-weighted scans of the lungs. Each scan had an in-plane resolution of 2.5mm $\times$ 2.5mm and a slice thickness of 3mm. The echo time was set at 60ms and the repetition time ranged from 2s to 4.4s, depending on the number of slices required to cover the lungs. Each patient underwent scanning with 6 different b-values (0, 50, 100, 200, 400, 600 sec/mm$^2$) in both axial and coronal planes with 6 gradient directions. Trace-weighted images were exported from the scanner A ROI was manually drawn for each case in the right lung at $b_0$ \cite{Afacan2016FetalAge}. 

The data set consists of 39 cases with different levels of misalignment between the different b-value image volumes. For each subject, we chose only one slice where the ROI in the right lung was labeled. The images were then cropped to a shape of $96 \times 96 $ and normalized by the 0.99 quantiles of the DWI image acquired without diffusion gradients (b-value=0 sec/mm$^2$). 

To ensure the reproducibility of our findings, we established two distinct, non-overlapping groups of 16 cases each for hyperparameter tuning. The composition of these groups was planned to encapsulate a wide array of gestational ages, thereby encompassing nearly the full breadth of ages present in our dataset. We conducted hyperparameter tuning for each group independently, as detailed in Section ~\ref{sec:hyperparam_tune}. The remaining 23 cases, that were left out in each group are designated as the test cases for each group. Our primary findings and analysis will be conducted on these specific cases.

In addition, we chose a sample of 10 cases for analysis. This sample included 5 cases exhibiting severe motion artifacts and another 5 with only minor motion artifacts. For each of these cases, we conducted a manual segmentation of one lung in the different b-value images.

\subsection{Baseline methods}
\label{sec:bsl_methods}
We compared our method to six baseline methods as follows:
\begin{enumerate}
    \item Quantifying IVIM parameters using the non-linear SLS-TRF approach \cite{gurney2018comparison, GradientMethodforLargeScale}, without utilizing any motion compensation.
    \item Registering all b-value images to the b=0 sec/mm$^2$ image using SyN registration \cite{AVANTS200826}, followed by quantifying IVIM parameters using the SLS-TRF algorithm.
    \item Registering all b-value images to the b=0 sec/mm$^2$ image using affine registration \cite{AVANTS200826}, followed by quantifying IVIM parameters using the SLS-TRF algorithm.
    \item Registering each b-value image to the previous image using SyN registration. For example, we register the b=50 sec/mm$^2$ to b=0 sec/mm$^2$ image and then register b=100 sec/mm$^2$ to the result.
    \item Iteratively quantifying IVIM parameters and registering each b-value image to the corresponding model image \cite{Kurugol2017Motion-robustEstimation}.
    \item Unsupervised VoxelMorph-based registration \cite{balakrishnan2019voxelmorph} of all b-value images to the b=0 sec/mm$^2$ image, followed by quantifying IVIM parameters using the SLS-TRF algorithm.
\end{enumerate}

\subsection{Hyper-parameters tuning}
\label{sec:hyperparam_tune}
For hyperparameter optimization, we implemented a grid search strategy, with a primary focus on determining the appropriate weights for the loss terms denoted as $\alpha_1$, $\alpha_2$, and $\alpha_3$. We selected the values for these hyperparameters as follows: $\alpha_1$ was varied within the range [0.5, 1, 5, 10], $\alpha_2$ within [0.015, 0.03], and $\alpha_3$ within [0.1, 0.8, 5].

The tuning process was conducted separately for two distinct groups, each comprising 16 cases, and was performed independently. The criterion used for selecting the optimal hyperparameters was based on the correlation between the IVIM parameter $f$ and gestational age during the canalicular stage of fetal development (GA $<$ 26 weeks).

\subsection{Lung Segmentation Evaluation}
\label{sec:lung_seg_ev}

We evaluated the anatomical registration accuracy of our IVIM-Morph in comparison to the different registration approaches outlined in Section~\ref{sec:bsl_methods} for cases with different levels of motion. These techniques were utilized to assess the alignment of images $S_i$ (for $i > 0$) with the reference image $S_0$. Our experiment involved 10 selected cases, each of which included manual segmentation of one lung. The effectiveness of these alignment methods was quantitatively assessed using the Dice score metric. This evaluation was conducted both prior to and following the application of registration.

\subsection{NCC loss contribution to the registration}
\label{sec:ncc_contribution}
We carried out an in-depth ablation study to thoroughly understand the impact and significance of the NCC loss on the registration process. To achieve this, we maintained constant values for certain parameters, setting $\alpha_1 = 1$ and $\alpha_2 = 0.015$. This controlled setup allowed us to isolate and examine the influence of the NCC loss more effectively.
We conducted this experiment by repeatedly executing the experiment outlined in Section ~\ref{sec:lung_seg_ev}, but with a key variation each time: we altered the value of $\alpha_3$ for each iteration. By systematically changing $\alpha_3$ while keeping the other parameters fixed, we were able to observe how variations in the NCC loss component affected the overall registration performance.
The series of experiments under varying $\alpha_3$ conditions were instrumental in gauging the sensitivity and responsiveness of our registration process to changes in the NCC loss.

\subsection{Clinical impact: Functional fetal lung maturity assessment}
\label{sec:clinical_imp}
We assessed the performance of our proposed method by examining its correlation with the GA and the perfusion fraction parameter ($f$) in the IVIM model. This parameter indirectly represents the proportion of the capillary network within the tissue. As established in prior research, the $f$ parameter exhibits a substantial increase with advancing gestational age in the fetal lung \cite{Ercolani2021IntraVoxelMaturation, korngut2022super}.
We conducted the analysis on the two group test cases (23 cases each).
% We conducted the analysis on three datasets: The first dataset comprised 39 cases, including 13 cases used for training; the second dataset included 26 cases that the network encountered for the first time; and the third dataset contained 13 cases marked as non-motion, which were used for training. We compared our proposed IVIM-Morph against baseline methods.% In  addition, we evaluated our approach when using a pre-trained IVIM-Morph as an initial solution and applied the IVIM-Morph as a self-supervised solver to compute the IVIM model parameters and deformation field for the current case (IVIM-Morph (solver)).
%Figure \ref{fig: clinical_exp} provides a schematic illustration of our proposed experiment.
For each case, we used IVIM-Morph to compute the IVIM parameter maps. Subsequently, we calculated the average parameter value in the lung for each case and evaluated the correlation between each parameter and GA separately for the canalicular and saccular phases, as suggested by \cite{korngut2022super}.

% \begin{figure*}[!t]
% \label{fig: clinical_exp}
% \centering
% \includegraphics[width=\textwidth]{figures/clinical_exp.pdf}
% \caption{Schematic illustration of our proposed clinical experiment. (a) Example of fetal lung DWI data with motion artifact. We feed each case separately to our IVIM-Morph solver to produce parameter maps and evaluate the averaged parameter value in the lung. (b) Correlation between each parameter and GA separately for the canalicular and saccular phases. The achieved correlations in the canalicular phase are $R^2_{D}=0.36$, $R^2_{D_p}=0.71$, and $R^2_{f}=0.55$.}
% \end{figure*}

\section{Results}
\subsection{Hyper-parameters tuning}
\label{sec:res-hp_tune}
The optimal hyperparameters for group 1 are: $\alpha_1 = 10, \alpha_2 = 0.015, \alpha_3 = 0.1$, and for group 2 are: $\alpha_1 = 0.5, \alpha_2 = 0.015, \alpha_3 = 0.8$.

\subsection{Lung Segmentation Evaluation}
\begin{figure}[!t]
\centering
\includegraphics[width=\columnwidth]{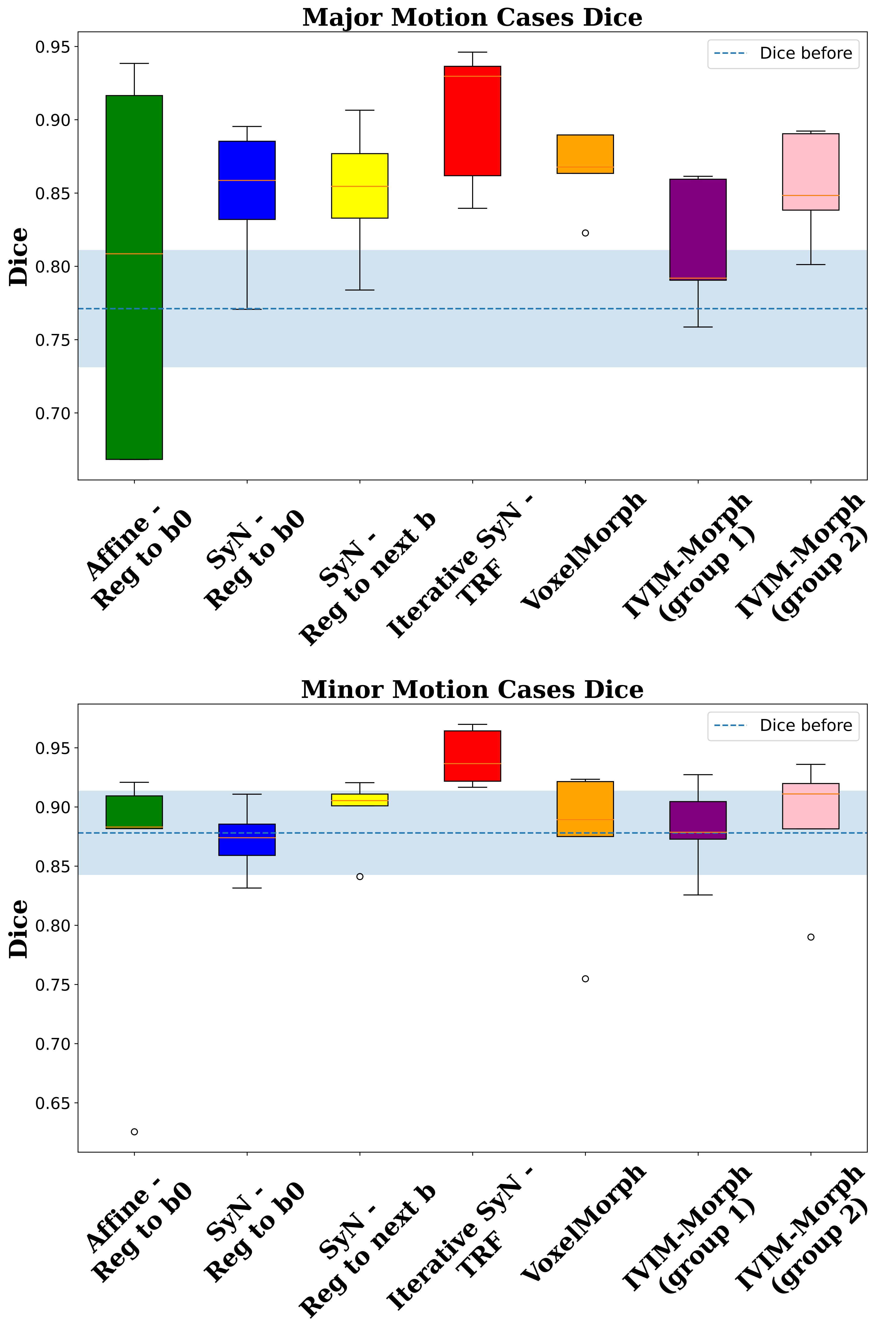}
\caption{Results from the Lung Segmentation Evaluation Experiment. The bar plot at the top illustrates the dice coefficients for cases with significant motion, while the bottom bar plot shows the dice coefficients for cases with minimal motion. In both plots, the dashed line and the shaded area indicate the average and standard deviation of the dice scores prior to the application of registration.}
\label{fig:subplot_dices}
\end{figure}

Lung segmentation evaluation results are presented in Figure~\ref{fig:subplot_dices}. The mean dice coefficient for each compared method is plotted as a boxplot, separately for the major and minor motion cases. We calculated the dice twice, one time using the optimal hyperparameters of group 1 and one time using the optimal hyperparameters of group 2. We also plotted the mean dice coefficient before applying registration. The mean dice before registration in the minor motion cases is $0.878\pm0.036$ and in the major motion cases is $0.771\pm0.040$, which is expected based on the cases' motion level. For cases involving major motion, IVIM-Morph succeeded in enhancing the dice coefficient achieving superior results for group 2 (dice = $0.854\pm0.038$) than group 1 (dice = $0.812\pm0.046$). Conversely, in scenarios with minor motion, IVIM-Morph, employing both sets of hyperparameters, consistently maintained a high dice coefficient.

\begin{figure*}[t!]
\centering
\includegraphics[width=0.98\textwidth]{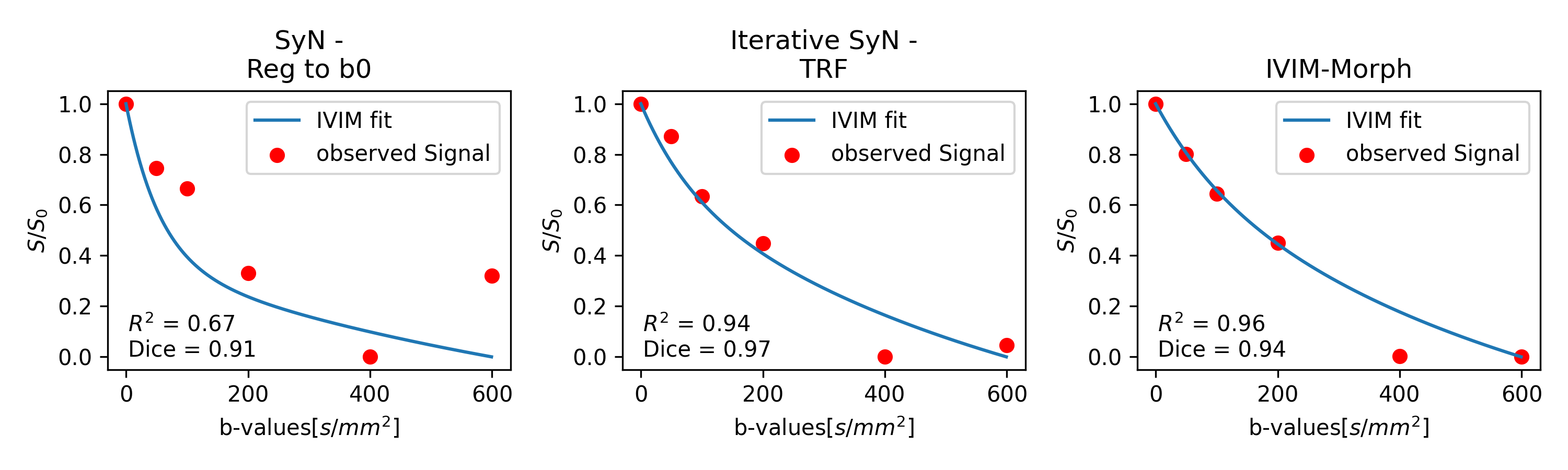}
\caption{Signal decay along the relaxation axis (b-value) after registration with SyN-Reg to b0, Iterative SyN-TRF, and IVIM-Morph. Although IVIM-Morph yielded a lower Dice score in this case compared to other methods, it better preserved the expected signal decay behavior, reflecting a more physically plausible registration.}
\label{fig:signal_vs_dice_exp}
\end{figure*}

Figure~\ref{fig:signal_vs_dice_exp} shows the signal decay along the b-value axis after registration using SyN-Reg to b0, Iterative SyN-TRF, and IVIM-Morph for a representative case. Although IVIM-Morph yielded a lower Dice score in this case, it better preserved the expected signal decay behavior, ensuring a more physically plausible registration. This balance between anatomical alignment and realistic signal decay is crucial for accurate IVIM parameter estimation, highlighting IVIM-Morph's effectiveness in maintaining signal integrity alongside reasonable anatomical registration.

\subsection{NCC loss contribution to the registration}
\begin{figure}[!t]
\centering
\includegraphics[width=\columnwidth]{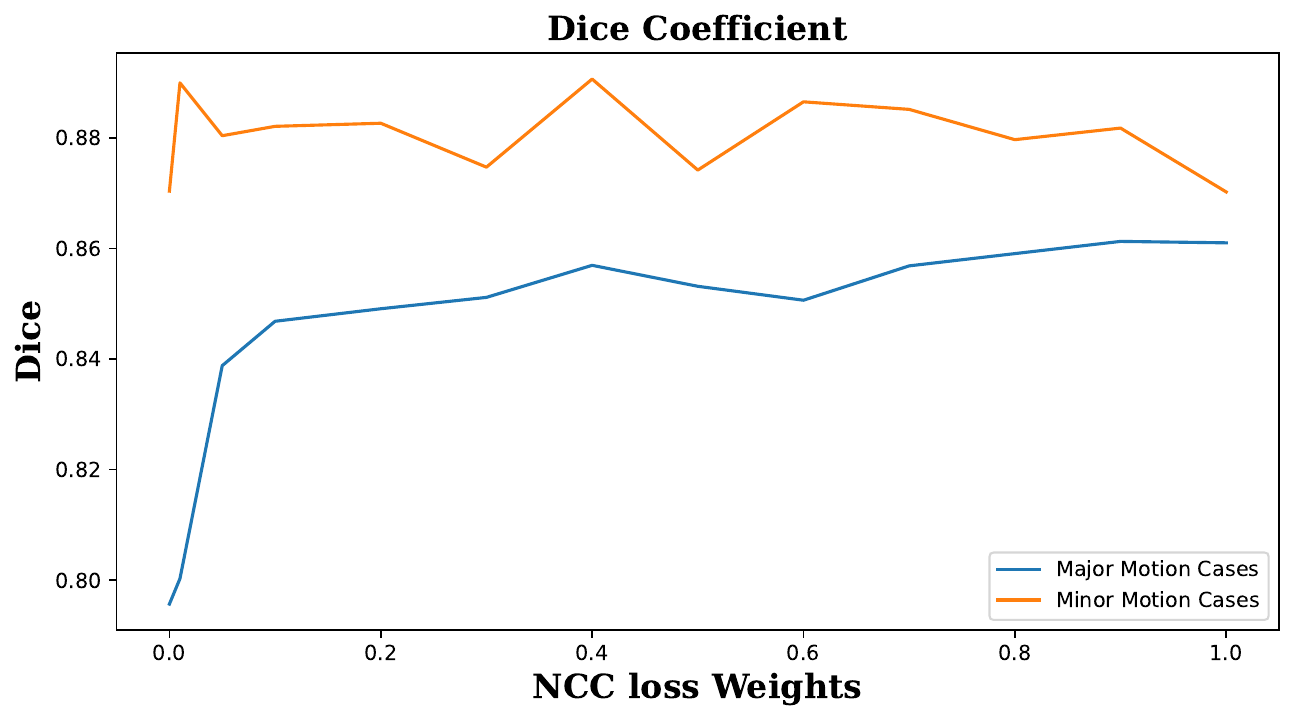}
\caption{Average dice coefficient computed for lung segmentation between $S_0$ and the deformed lung segmentation in $S_i$ (for $i > 0$), utilizing IVIM-Morph in 10 specifically chosen cases with varying $\alpha_3$ values. The blue line indicates the cases characterized by major motion, whereas the orange line corresponds to the cases with minor motion.}
\label{fig:ncc_contribution_exp}
\end{figure}

Figure ~\ref{fig:ncc_contribution_exp} displays the outcomes of the experiment investigating the impact of NCC loss. In scenarios with minor motion, the choice of $\alpha_3$ value seems to have a negligible effect on the dice coefficient achieved post-application of IVIM-Morph. Conversely, in cases of major motion, a lower weighting on NCC loss is observed to yield suboptimal dice scores. This finding highlights the increased importance of NCC loss in the optimization process, particularly in instances where major motion artifacts are present.

\begin{table}[!t]
\small
\caption{\label{Table:times} Methods running times}
\centering

\begin{tabular}{|c|c|c|}

\hline
\textbf{Method}        & \textbf{Time (s)} & \textbf{Machine} \\ \hline
SLS-TRF                      & $57.93 \pm 1.26$ & CPU \\ \hline
Affine - Reg to b0           & $53.32 \pm 3.09$ & CPU \\ \hline
SyN - Reg to b0              & $54.04 \pm 2.14$ & CPU \\ \hline
RSyN - Reg to next b         & $53.11\pm  2.29$ & CPU \\ \hline
Iterative SyN-TRF            & $261.16 \pm 67.61$ & CPU     \\ \hline
VoxelMorph + SLS - TRF       & $84.738 \pm 7.23$ & CPU+GPU \\ \hline
IVIM-Morph                   & $52.69 \pm 1.90$  & GPU \\ \hline

\end{tabular}
\end{table}

% The deformation field prediction accuracy results are presented in Figure~\ref{fig: std_exp}. The Dice coefficient is plotted as a function of the simulated motion level after applying IVIM-Morph (blue) compared to the baseline dice before motion correction. IVIM-Morph effectively compensated for motion at different levels while avoiding motion correction in cases without motion (motion level = 0). The Dice coefficient decreased as expected for higher levels of motion, which presented a more complex task. However, it is important to note that synthetic motion does not accurately reflect fetal motion as it does not follow any motion model. Additionally, for high motion levels, the synthetic motion may be unrealistic.
% The slight decrease in the Dice coefficient at motion level zero may be due to the compensation of IVIM-Morph for acquisition noise while predicting the IVIM model parameters.

\subsection{Clinical impact: Functional fetal lung maturity assessment}

\begin{figure*}[!t]
\centering
\includegraphics[width=1\textwidth]{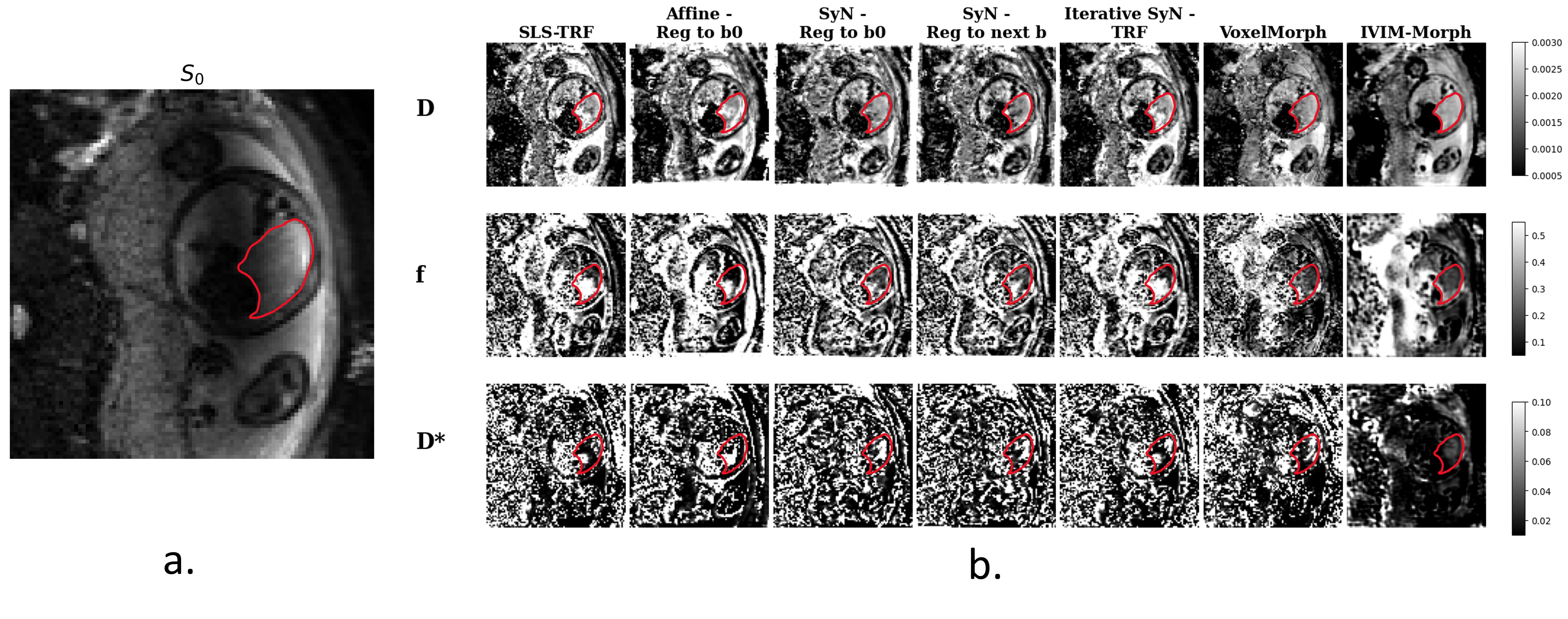}
\caption{Comparison of IVIM parameter estimation methods for a single subject. (a) Baseline $S_0$ image. (b) IVIM parameter maps ($D_p$, $D_t$, and $F$) estimated using different approaches: SLS-TRF, Affine registration to $S(b_0)$, SyN registration to $S(b_0)$, SyN registration to the previous b-value image, Iterative SyN-TRF, VoxelMorph, and IVIM-Morph. The contours of one lung are shown for reference.}
\label{fig: params_maps}
\end{figure*}

Figure~\ref{fig: params_maps} shows representative IVIM parameter maps generated by each method for a case with motion and for a case without motion. The IVIM-Morph method produced smoother parameter maps compared to the other methods. This can be attributed to the use of a CNN in the qDWI sub-network which leverages spatial correlations to estimate IVIM parameters more accurately and results in smoother parameter maps. In contrast, the other methods rely on traditional optimization and registration techniques, which may result in less accurate and noisier parameter maps.

% \begin{figure*}[!t]
% \centering
% \includegraphics[width=0.9\textwidth]{figures/param_with_wo_motion.pdf}
% \caption{Example of the parameter maps ($D_p, D_t$ and $f$) estimated using different methods, including SLS-TRF, registration to $S(b_0)$, registration to the previous b-value image, Iterative SyN-TRF and IVIM-Morph. (a) Case with motion. (b) Case without motion.}
% \label{fig: params_maps}
% \end{figure*}

\begin{figure}[!t]
\centering
\includegraphics[width=0.47\textwidth]{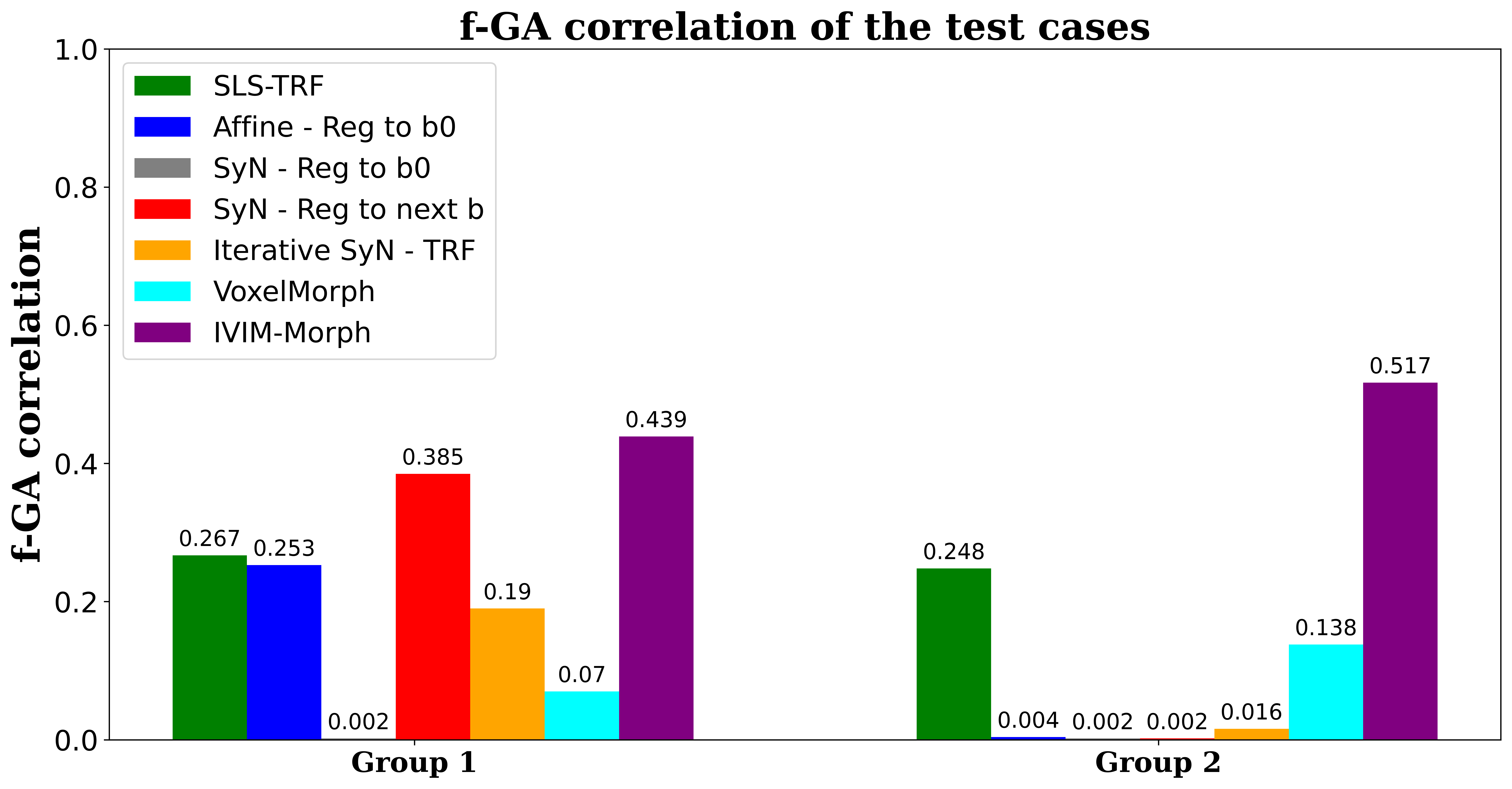}
\caption{The correlations between f and the GA in the canalicular stage are calculated in two datasets: group 1 test cases and group 2 test cases.}
\label{fig:test_cases_corrs}
\end{figure}

Figure~\ref{fig:test_cases_corrs} presents the correlation analysis between the IVIM parameters as computed by each method and the GA. Our IVIM-Morph approach outperformed the other methods, achieving the highest correlation coefficient ($0.44$ for group 1 and $0.52$ for group 2) for the $f$ parameter in the canalicular phase. 
Figure ~\ref{fig:test_cases_corrs} displays correlations derived from two test groups, where each group's test cases include those from the opposing original group (the 16 cases for the hyperparameter tune). Average IVIM parameters were computed in the ROI for each case across all evaluated methods, utilizing the best hyperparameters for each group as detailed in \ref{sec:res-hp_tune} for the IVIM-Morph calculations. Notably, IVIM-Morph demonstrates greater consistency in terms of correlation between the two groups, unlike other methods (except TRF-SLS and Syn - Reg to $b_0$), which showed varying correlations across the groups.

Supplementary material includes supplementary results and tables that summarize the correlations between different IVIM model parameters and GA.

\section{Discussion and Conclusions}
Accurately assessing IVIM parameters while addressing fetal movement is crucial for obtaining precise quantitative imaging biomarkers related to fetal lung development. In this study, we introduced IVIM-Morph, a self-supervised deep neural network approach designed for simultaneous motion compensation and quantitative DWI analysis using the IVIM model. Our method surpassed baseline approaches that consider motion and estimate IVIM model parameters, notably enhancing the correlation between the perfusion fraction parameter of the IVIM model ($f$) and gestational age (GA). 

Our segmentation experiment results, shown in Figure \ref{fig:subplot_dices}, indicate that the iterative SyN-TRF method achieves improved alignment, as evidenced by a higher Dice coefficient. However, this alignment primarily impacts boundary regions rather than the signal within the organ itself, which is critical for accurate IVIM parameter estimation across the entire organ region. As a result, while SyN-TRF improves segmentation at the boundaries, it does not ensure realistic signal decay along the full b-value axis, leading to inaccurate estimates of the IVIM parameters and a weak correlation between the pseudo-diffusion fraction parameter $f$ and gestational age (GA).

In contrast, our IVIM-Morph approach balances precise boundary alignment with realistic signal decay across the entire b-value range, enhancing both segmentation accuracy and parameter reliability. This approach maintains segmentation performance comparable to SyN-TRF while preserving physiologically meaningful signal behavior within the organ region, resulting in a significantly improved correlation between $f$ and GA. Thus, IVIM-Morph provides a more robust framework for IVIM parameter estimation, bridging the gap between accurate segmentation and reliable signal-based analysis within the organ.

An important consideration when training neural networks for IVIM parameter estimation guided by the signal decay model is the potential for circularity, particularly if networks are later evaluated based on their correlation with biological measures like gestational age (GA). Our network design specifically avoids this issue by excluding any GA-related information during training. Instead, the network is trained solely on the signal decay model, focusing on accurately capturing diffusion characteristics without any knowledge of GA. This approach ensures that the estimation of IVIM parameters, such as the pseudo-diffusion fraction $f$, is derived purely from signal properties, independent of the biological measurement assessed later. Consequently, the observed correlation between $f$ and GA is not an artifact of circularity but an independent validation of the network's ability to capture meaningful diffusion-related information relevant to gestational development.

It's crucial to emphasize that, in contrast to previously proposed methods primarily addressing motion between slices within a single volume in anatomical fetal imaging, our IVIM-Morph focuses on addressing motion between volumes in quantitative DWI acquisitions that encompass multiple volumes. Additionally, in its current configuration, IVIM-Morph operates under the assumption of a single trace-weighted image per b-value, which is automatically generated by aggregating the various b-vector images into a single scalar map. Future enhancements may involve accommodating motion between the distinct b-vector images employed in the computation of the trace-weighted b-value image.

While our primary emphasis has been on the quantitative analysis of fetal lung DWI data, it's worth noting that the proposed approach holds potential applicability in various other quantitative DWI analysis domains that grapple with motion-related challenges. For instance, applications such as the detection and staging of liver fibrosis through IVIM analysis of abdominal DWI data \cite{ye2020value}, the assessment of non-alcoholic fatty liver disease \cite{guiu2012intravoxel}, and the identification of diffuse renal pathologies \cite{caroli2018diffusion} can all derive benefits from our approach by effectively accounting for motion induced by processes like respiration between different b-value volumes.

In conclusion, our study has showcased the clinical promise of evaluating functional fetal lung maturation non-invasively from fetal DWI data. our IVIM-Morph stands out as a means to markedly enhance the precision of non-invasive, quantitative fetal lung maturity assessment. Moreover, our proposed method can be readily extended to other clinical scenarios necessitating motion correction in the computation of quantitative MRI biomarkers. Collectively, our findings point to IVIM-Morph as a valuable asset in advancing fetal MRI, enhancing fetal health monitoring, and informing clinical decision-making.

\section*{Acknowledgments}
This research was supported in part by the United States-Israel Binational Science Foundation (BSF), Jerusalem, Israel under award number 2019056, and by the National Institutes of Health (NIH) under award numbers R01 LM013608, R01 EB019483, R01 NS124212, and R01 NS121657. The content is solely the responsibility of the authors and does not necessarily represent the official views of the National Institute of Health.

\section*{Declaration of Generative AI}
During the preparation of this work, the author(s) used ChatGPT in order to improve readability. After using this tool/service, the author(s) reviewed and edited the content as needed and take(s) full responsibility for the content of the publication.

% Please ensure that every reference cited in the text is also present in
% the reference list (and vice versa).

% \section*{\itshape Reference style}

% Text: All citations in the text should refer to:
% \begin{enumerate}
% \item Single author: the author's name (without initials, unless there
% is ambiguity) and the year of publication;
% \item Two authors: both authors' names and the year of publication;
% \item Three or more authors: first author's name followed by `et al.'
% and the year of publication.
% \end{enumerate}
% Citations may be made directly (or parenthetically). Groups of
% references should be listed first alphabetically, then chronologically.

%%Harvard
\bibliographystyle{model2-names.bst}\biboptions{authoryear}
\bibliography{refs}

% \section*{Supplementary Material}

% Supplementary material that may be helpful in the review process should
% be prepared and provided as a separate electronic file. That file can
% then be transformed into PDF format and submitted along with the
% manuscript and graphic files to the appropriate editorial office.

\end{document}

\maketitle

\begin{figure}[!ht]
\centering
\includegraphics[width=1.05\textwidth]{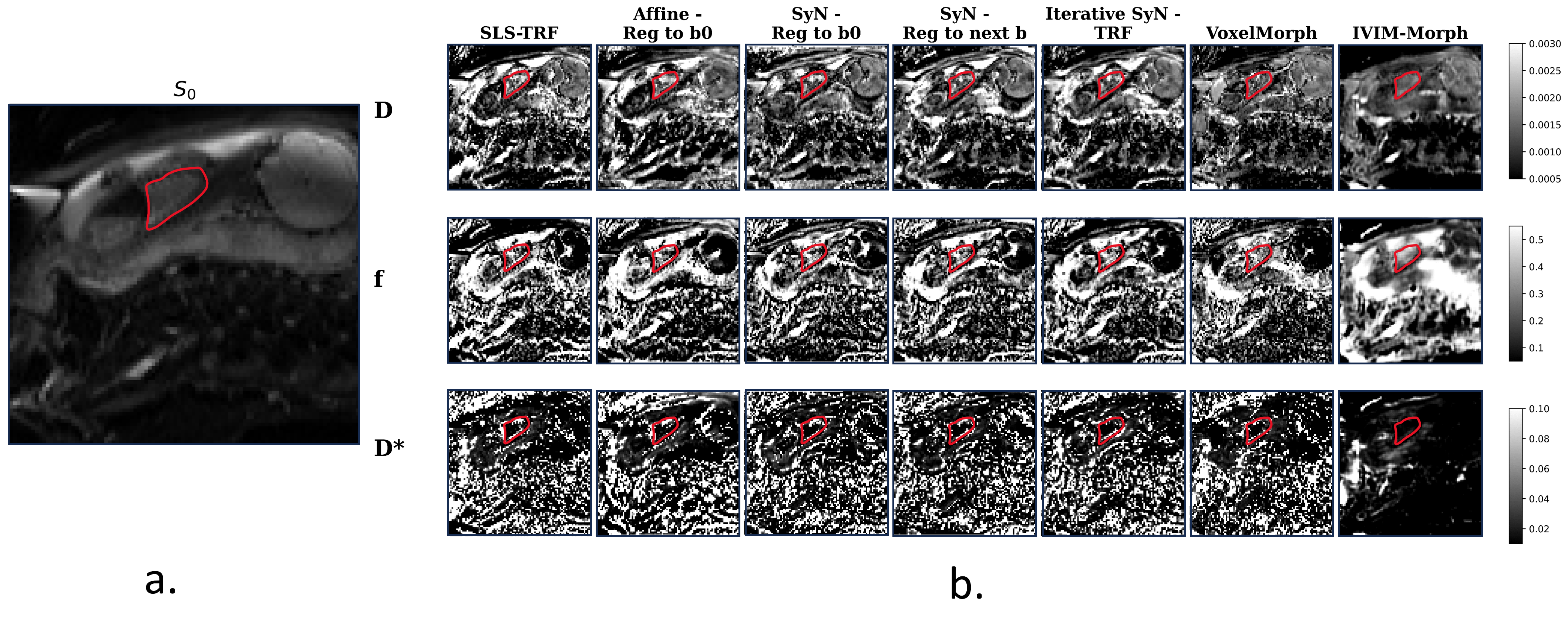}
\caption{a.) Example of baseline $S_0$ image .b) IVIM parameter maps($D_p, D_t$ and $F$) of the same subject estimated using different methods, including SLS-TRF, registration to $S(b_0)$ using Affine registration, registration to $S(b_0)$ using SyN registration, registration to the previous b-value image using SyN registration, Iterative SyN-TRF, VoxelMorph and IVIM-Morph. The contours of one lung are displayed}
\label{fig: params_maps}
\end{figure}
\newpage
\begin{table}[!ht]
    \caption{Group 1 tests cases IVIM parameters correlation with GA in the canicular and saccular stages}
    \centering
    \begin{tabular}{|l|l|l|l|l|l|l|}
    \hline
        ~ & Dt\_can & Dt\_sac & Dp\_can & Dp\_sac & Fp\_can & Fp\_sac \\ \hline
        SLS-TRF & 0.111 & 0.118 & 0.279 & 0.008 & 0.267 & 0.107 \\ \hline
        Affine-TRF reg to b0 & 0.022 & 0.089 & 0.052 & 0.004 & 0.253 & 0.202 \\ \hline
        SyN-TRF reg to b0 & 0.145 & 0.000 & 0.021 & 0.002 & 0.002 & 0.000 \\ \hline
        SyN-TRF reg to next b & 0.297 & 0.018 & 0.023 & 0.001 & 0.385 & 0.009 \\ \hline
        Iterative SyN-TRF & 0.203 & 0.124 & 0.004 & 0.001 & 0.190 & 0.048 \\ \hline
        \textbf{IVIM-Morph} & 0.075 & 0.417 & 0.286 & 0.040 & \textbf{0.439} & 0.125 \\ \hline
        VoxelMorph & 0.074 & 0.101 & 0.254 & 0.026 & 0.070 & 0.121 \\ \hline
    \end{tabular}
\end{table}

\begin{table}[!ht]
    \caption{Group 2 tests cases IVIM parameters correlation with GA in the canicular and saccular stages}
    \centering
    \begin{tabular}{|l|l|l|l|l|l|l|}
        \hline

        ~ & Dt\_can & Dt\_sac & Dp\_can & Dp\_sac & Fp\_can & Fp\_sac \\ \hline
        SLS-TRF & 0.007 & 0.000 & 0.299 & 0.058 & 0.249 & 0.027 \\ \hline
        Affine-TRF reg to b0 & 0.007 & 0.042 & 0.360 & 0.216 & 0.004 & 0.007 \\ \hline
         SyN-TRF reg to b0 & 0.095 & 0.007 & 0.478 & 0.129 & 0.003 & 0.017 \\ \hline
         SyN-TRF reg to next b & 0.023 & 0.000 & 0.445 & 0.056 & 0.002 & 0.021 \\ \hline
        Iterative SyN-TRF & 0.000 & 0.002 & 0.420 & 0.063 & 0.016 & 0.001 \\ \hline
         \textbf{IVIM-Morph} & 0.400 & 0.008 & 0.425 & 0.013 & \textbf{0.517} & 0.042 \\ \hline
        VoxelMorph & 0.027 & 0.132 & 0.069 & 0.068 & 0.139 & 0.121 \\ \hline
    \end{tabular}
\end{table}

\begin{figure}[!t]
\centering
\includegraphics[width=1\textwidth]{all_cases_corrs_group1_test_cases_final.png}
\caption{Group 1: Scatter plots depict the average f parameter values in the fetal lung across all methods evaluated. Separate linear fitting with GA was performed for both the canalicular and saccular stages.}
\end{figure}

\begin{figure}[!t]
\centering
\includegraphics[width=1\textwidth]{all_cases_corrs_group2_test_cases_final.png}
\caption{Group 2: scatter plots depict the average f parameter values in the fetal lung across all methods evaluated. Separate linear fitting with GA was performed for both the canalicular and saccular stages.}
\end{figure}